\documentclass[useAMS]{mn2e}

\usepackage{times}
\usepackage{amssymb}
\usepackage{rotating}
\usepackage{graphicx}
\usepackage{longtable}
\usepackage{array}
\usepackage{multirow}
\usepackage{tabularx}
\usepackage{caption}

\title{Evolution of the hot flow of MAXI J1543-564}
\author[S. Rapisarda, A. Ingram, and M. van der Klis]{S. Rapisarda, A. Ingram, and M. van der Klis}

\date{Submitted to MNRAS}
\pagerange{\pageref{firstpage}--\pageref{lastpage}} \pubyear{2013}

\begin{document}



\maketitle
\topmargin = -0.5cm
\label{firstpage}

\begin{abstract}

We present a spectral and timing analysis of the black hole candidate MAXI J1543-564 during its 2011 outburst. As shown in previous work, the source follows the standard evolution of a black hole outburst. During the rising phase of the outburst we detect an abrupt change in timing behavior associated with the occurrence of a type-B quasi-periodic oscillation (QPO). This QPO and the simultaneously detected radio emission mark the transition between hard and soft intermediate state. We fit power spectra from the rising phase of the outburst using the recently proposed model \textsc{propfluc}. This assumes a truncated disc / hot inner flow geometry, with mass accretion rate fluctuations propagating through a precessing inner flow. We link the \textsc{propfluc} physical parameters to the phenomenological multi-Lorentzian fit parameters. The physical parameter dominating the QPO frequency is the truncation radius, while broad band noise characteristics are also influenced by the radial surface density and emissivity profiles of the flow. In the outburst rise we found that the truncation radius decreases from $r_o \sim 24$ to $10 R_g$, and the  surface density increases faster than the mass accretion rate, as previously reported for XTE J1550-564. Two soft intermediate state observations could not be fitted with \textsc{propfluc}, and we suggest that they are coincident with the ejection of material from the inner regions of the flow in a jet or accretion of these regions into the BH horizon, explaining the drop in QPO frequency and suppression of broad band variability preferentially at high energy bands coincident with a radio flare.
\end{abstract}


\section{Introduction}
\label{sec:int}

Transient black hole binaries (BHBs) display outbursts exhibiting several states, characterized by both spectral and timing properties (e.g. Belloni et al. 2005; Remillard \& McClintock 2006; Belloni 2010; Gilfanov 2010). During the outburst, sources typically follow a 'q' shaped, anti-clockwise track on a plot of X-ray flux versus spectral hardness ratio (hardness-intensity diagram: HID), with the quiescent state occupying the bottom right corner. The initial transition from hard (LHS) to soft (HSS), via intermediate states, occurs when the power low component of the spectrum is observed to soften (photon index $\Gamma \sim$ 1.7--2.4) and a disc blackbody component (peaking in soft X-rays) becomes increasingly prominent. A power spectral analysis of the rapid variability reveals a quasi-periodic oscillation (QPO), which shows up as narrow harmonically related peaks, superimposed on broad band continuum noise. The QPO fundamental frequency is observed to increase from $\sim$ 0.1--10 Hz during the transition from the hard state, after which the X-ray emission becomes very stable in the soft state. Power spectral evolution correlates tightly with spectral evolution, with all the characteristic frequencies increasing with spectral hardness (e.g. Wijnands \& van der Klis 1998; Psaltis, Belloni \& van der Klis 1999; Homan et al. 2001). QPOs observed coincident with broad band noise are defined as Type-C QPOs (Remillard et al. 2002; Casella et al. 2005). Type-B QPOs (Wijnands, Homan \& van der Klis 1999), typically with a frequency of $\sim$ 6--10 Hz, are observed in the intermediate state when the broad band noise suddenly disappears. These features quickly evolve into Type-A QPOs (Wijnands, Homan \& van der Klis 1999), which are broader and weaker. Since the sudden suppression of the broad band noise hints a large physical change in the system, intermediate state observations displaying Type-C QPOs are classified as hard intermediate state (HIMS) and those displaying Type-A or B QPOs as soft intermediate state (SIMS). Additionally, a large radio flare, indicative of a jet ejection event, is often observed to be coincident with the onset of the SIMS (Fender, Belloni \& Gallo 2004), although this is not always exact (Fender, Belloni, \& Gallo 2005).\\ 
The spectral and timing properties of BHBs can be described by \textit{the truncated disc model} (e.g. Esin, McClintock \& Narayan 1997; Done, Gierli\'nski \& Kubota 2007) where an optically thick, geometrically thin accretion disc which produces the multi-temperature blackbody spectral component (Shakura \& Sunyaev 1973) truncates at some radius, $r_o$, larger than the innermost stable circular orbit (ISCO). In the region between this truncation radius $r_o$ and an inner radius $r_i$ ($r_o > r_i > r_{ISCO}$), accretion takes place via a hot, optically thin, geometrically thick accretion flow (hereafter inner flow). Compton up-scattering of cool disc photons by hot electrons in the flow produces the power low spectral component (Thorne \& Price 1975; Sunyaev \& Truemper 1979). In the hard state $r_o$ is large ($\sim 60 R_g$, where $R_g=GM/c^2$ is a gravitational radius), so only a small fraction of the disc photons illuminates the flow, giving rise to a weak direct disc component and hard power law emission. As the average mass accretion rate increases during the outburst, $r_o$ decreases, so more direct disc emission is seen and a greater luminosity of disc photons cool the flow, resulting in softer power law emission. When $r_o$ reaches the ISCO, the direct disc emission completely dominates the spectrum and the transition to the soft state is complete.\\
This scenario is the framework of the propagating fluctuations model \textsc{propfluc} (Ingram \& Done 2011, 2012, hereafter ID11, ID12; Ingram \& Van der Klis 2013, IK13), a model that can reproduce power density spectra by combining the effects of the propagation of mass accretion rate fluctuations in the inner flow (Lyubarskii 1997; Arevalo \& Uttley 2006), responsible for generating the broad band noise, with solid-body Lense-Thirring (LT) precession of this flow (Fragile et al. 2007; Ingram, Done \& Fragile 2009), producing QPOs. Mass accretion rate fluctuations are generated throughout the inner flow, with the contribution to the rms variability from each region peaking at the local viscous frequency (e.g. Lyubarskii 1997; Churazov, Gilfanov \& Revnivtsev 2001; Arevalo \& Uttley 2006), thus the fast variability originates from the inner regions and the slow variability from the outer regions. As material is accreted, fluctuations propagate inwards, modulating the faster variability generated in the inner regions. Emission is thus highly correlated from all regions of the flow, giving rise to the observed linear rms-flux relation (Uttley \& McHardy 2001; Uttley, Vaughan \& McHardy 2005).\\
In this paper we present a spectral and timing analysis of the source MAXI J1543-564 during its 2011 outburst. The source, discovered by MAXI/GSC (the Gas Slit Camera of the Monitor of ALL-sky X-ray Image; Matsuoka et al. 2009) on May 08 2011 (Negoro et al. 2011), was first analyzed by Stiele et al. (2012). Their analysis showed that the outburst evolution follows the usual BHBs behavior, the exponential flux decay is interrupted by several flares, and during the transition from LHS and HSS a type-C QPO is observed. Looking at other wavelengths, Miller-Jones et al. (2011) report the detection of radio emission at MJD 55695.73. In this work, we analyze the spectral and timing properties of the source in different energy bands and we use the power density spectra of the rising phase of the outburst to systematically explore for the first time the capabilities of \textsc{propfluc}.\\ 

\section[]{Observations and data analysis}
\label{sec:obs}

We analyzed data from the RXTE Proportional Counter Array (PCA; Jahoda et al. 1996) using 99 pointed observations collected between 10 May and 30 September 2011. Each observation consisted of between 300--4750 s of useful data. \\
We used Standard 2 mode data (16 s time resolution) to calculate a hard color (HC) as the 16.0--20.0 / 2.0--6.0 keV count rate ratio and define the intensity as the count rate in the 2.0--20.0 keV band. All the observations were background subtracted and all count rates were normalized by the corresponding Crab values closest in time to the observations.\\
We used the $\sim$ 125 $\mu$s time resolution Event mode and the $\sim$ 1 $\mu$s time resolution Good-Xenon mode data for Fourier timing analysis. We constructed Leahy-normalized power spectra using 128 s data segments and 1/8192 s time bins to obtain a frequency resolution of 1/128 Hz and a Nyquist frequency of 4096 Hz. After averaging these power spectra per observation, we subtracted the Poisson noise using the method developed by Klein-Wolt et al. (2004), based on the expression of Zhang et al. (1995), and renormalized the spectra to power density $P_{\nu}$ in units of $(rms / mean)^2$ / Hz. In this normalization the fractional rms of a variability component is directly proportional to the square root of its integrated power density: $rms = 100 \sqrt{ \int_{0}^{\infty} P_{\nu} d\nu} $ \%. No background or dead-time corrections were made in computing the power spectra. This procedure was performed in 4 different energy bands: 2.87--4.90 keV (band 1), 4.90--9.81 keV (band 2), 9.81--20.20 keV (band 3), and the full 2.87�-20.20 keV (band 0). The power spectra were fitted using a multi-Lorentzian function in which each Lorentzian contributing to the fit function is specified by a characteristic frequency $\nu_{max}=\sqrt{{\nu_0}^2+(FWHM/2)^2}$ (Belloni, Psaltis \& van der Klis 2002) and a quality factor $Q=\nu_{0}/FWHM$, where FWHM is the full width at half maximum and $\nu_0$ is the centroid frequency of the Lorentzian. All the power spectra shown in this paper were plotted using the power times frequency representation ($\nu P_{\nu}$), in order to visualize $\nu_{max}$ as the frequency where the Lorentzian's maximum occurs. \\

\section[]{Results}
\label{sec:res}

\subsection{Light curve}

\begin{figure} 
\center
\includegraphics[scale=0.4,angle=270]{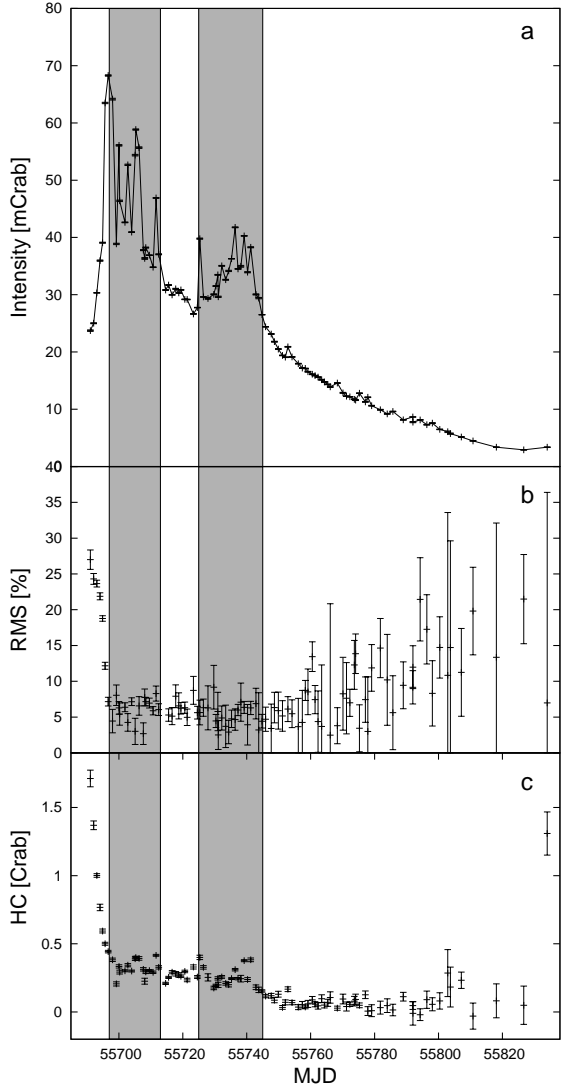}
\caption{a) Intensity [mCrab], b) rms [\%], c) and hard color [Crab] versus time for the 99 pointed observations. The grey rectangular areas indicate 5 time intervals characterized by different long-term luminosity variability. Data points are plotted with 1$\sigma$ error bars.}
\label{fig:tid}
\end{figure}

The light curve of the source is shown in Fig. \ref{fig:tid}$a$, where the 2-20 keV intensity is plotted versus time (MJD) for each pointed observation. \\
We subdivided the evolution of the outburst in 5 parts. In the first part of the outburst (MJD 55691--55696 ) the source rises to maximum intensity ($\sim$ 68 mCrab) in 5 days from the beginning of the RXTE observations. The second part (first grey area, MJD 55696--55713) is characterized by an intensity decay that is not smooth, but interrupted by 4 additional peaks with intensities between $\sim$ 47 and $\sim$ 58 mCrab. The third part (MJD 55713--55725, between the two grey areas) does not show any intensity peak but only a gradual decay. The following period (MJD 55725--55744, second grey area) is characterized by a broad maximum and several additional intensity peaks (between  $\sim$ 34 and $\sim$ 42 mCrab) less luminous compared to those of the first grey area. Finally, the last part (MJD 55744--55834) consists of a relatively smooth decay until the end of the observations. \\

\subsection{Color diagrams}

\begin{figure} 
\center
\includegraphics[scale=0.4,angle=270]{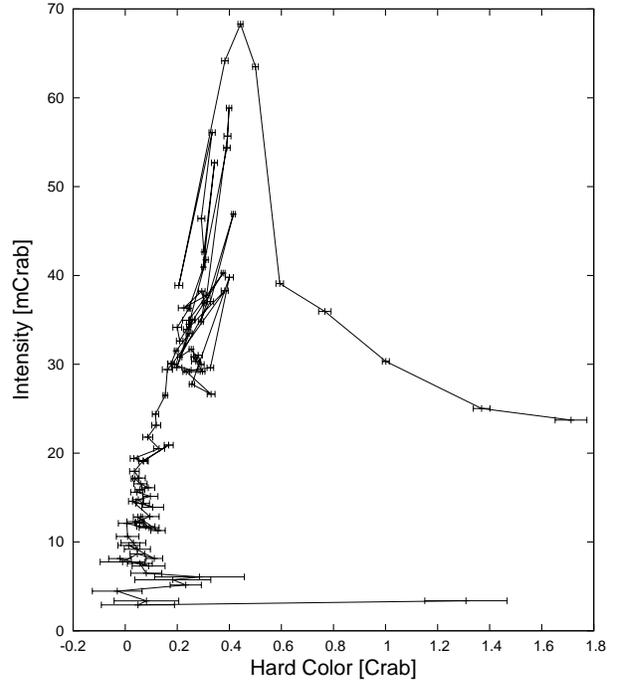}
\caption{Hard color versus Intensity normalized to the Crab. Points represent average intensity and hard color for each observation. 1$\sigma$ error bars are plotted for the hard color.}
\label{fig:hid}
\end{figure}

Fig. \ref{fig:hid} shows the hardness��-intensity diagram (HID), where the average intensity of each observation is plotted versus the HC. The source follows a counterclockwise path, starting and ending in the right (hard) part of the diagram at different luminosities. This is the usual behavior observed for black hole outbursts.\\
In order to better follow the spectral evolution of the source along the outburst, we also plotted in Fig. \ref{fig:tid}$c$ the HC versus time.\\
In the first observation the source is harder than Crab (HC = 1.71) and in the following 6 observations softens continuously, while at the same time its intensity increases from $\sim$ 24 to $\sim$ 68 mCrab. For the remaining observations the source remains in the soft part of the HID (HC$\le$0.5) except for the very last observation, where it goes back to a color harder than Crab (HC = 1.31$\pm$0.16).\\
As can be noted in Fig. \ref{fig:tid}$c$, the transitions between hard and soft spectrum happen on short time scales ($\sim$ 10 days) compared to the time spent by the source in the soft state ($\sim$ 125 days). However, while the initial transition from hard to soft state is simultaneous with a quick change in intensity (+ 188$\%$), the final transition (last observation) from soft to hard spectrum is characterized by a fractional intensity change of only + 16$\%$, i.e. increasing when the source gets harder.\\

\subsection{Time variability}

The 1/129--10 Hz rms values as computed from the power spectra in band 0 are reported in Fig. \ref{fig:tid}$b$. The first 5 observations, during which the source rapidly becomes softer and brighter are characterized by rms values of $\sim$ 19--27$\%$. In the remaining observations the rms values are between $\sim$ 2$\%$ and $\sim$ 10$\%$ with few exceptions.\\
Integrated rms is systematically higher for higher energies. From the beginning of the observations, as the intensity increases, integrated rms decreases independently from photon energy, but the rms decrease trend is different between energy bands. In order to better show these differences, we plotted in Fig. \ref{fig:rmsall} the total fractional rms of the first 7 observations for all the energy bands. Band 1 (red) shows a smooth and continuous rms decrease with time, while in band 2 (green) and band 3 (blue), the rms decrease is characterized by a ''jump'' between observations \#5 and \#6 ($\Delta$rms $\sim$ --9$\%$ in band 2, $\Delta$rms $\sim$ --11$\%$ in band 3). Observation \#6 is also characterized by the detection of radio emission, indicated by the orange arrow.\\

\subsubsection{QPOs and broad band features}

\begin{figure} 
\center
\includegraphics[scale=0.4,trim=1cm 3cm 2cm 1cm,clip,angle=270]{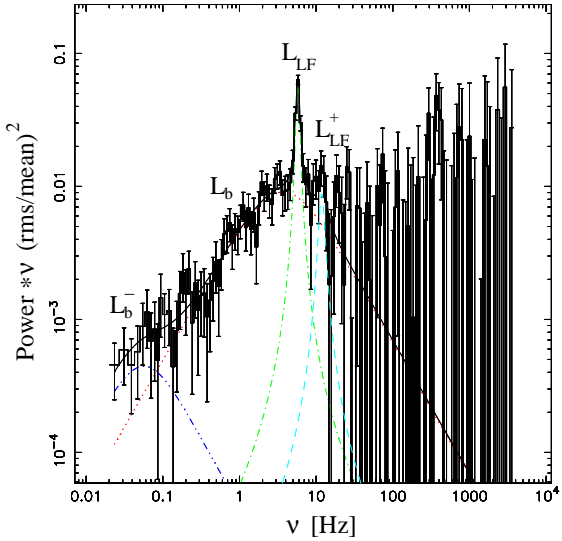}
\caption{Multi-Lorentzian fit of the fifth power spectrum. Four main components were identified: a main QPO $L_{LF}$, its harmonic $L_{LF}^{+}$, a broad band noise component $L_b$, and another broad component at lower frequency $L_b^{-}$.}
\label{fig:psex}
\end{figure}

\begin{figure} 
\center
\includegraphics[scale=0.4,trim=1cm 3cm 1cm 0cm,clip,angle=270]{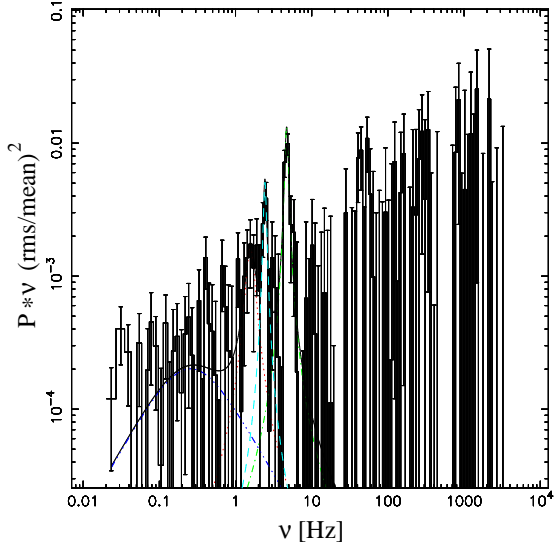}
\caption{Lorentzian fit of observation $\#$7 showing a type-B QPO.}
\label{fig:typeb}
\end{figure}

\begin{figure*} 
\center
\includegraphics[scale=0.35,trim=0cm 0cm 0cm 0cm,angle=270]{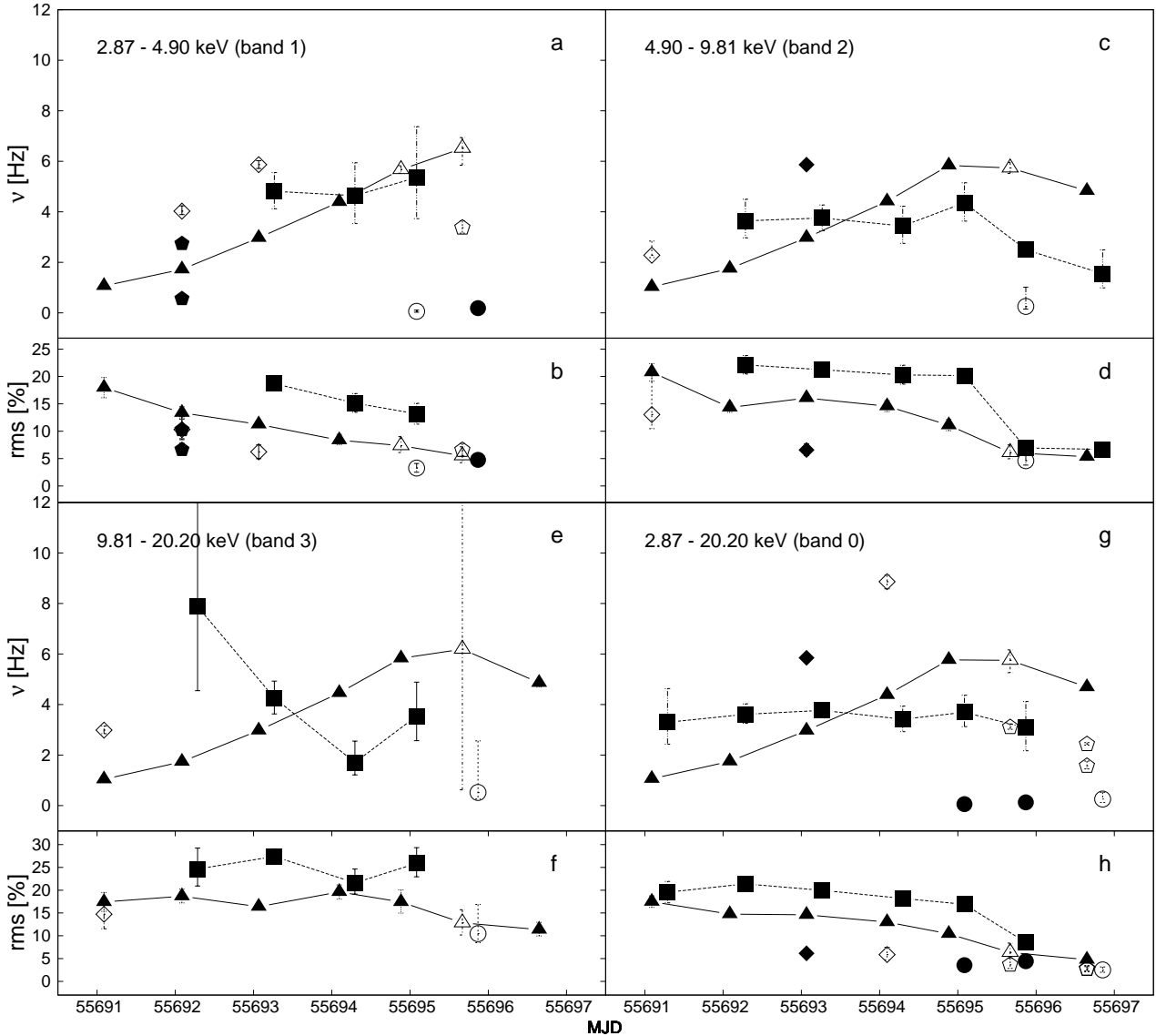}
\caption{Characteristic frequency and rms of $L_{LF}$ (triangles), $L_{LF}^+$ (diamonds), $L_b$ (squares), $L_b^{-}$ (circles), and other significant unidentified components (pentagons) fitted in the first 7 observations in all the energy bands ($L_b$ and $L_b^{-}$ have been shifted slightly to the right with respect to the original position for clear reading). Open symbols indicate components with significance between 2 and 3 $\sigma$ while full symbols stand for $\sigma > 3$ significant components. All values are plotted with 1$\sigma$ error bars.} 
\label{fig:fmax}
\end{figure*}

\begin{figure} 
\center
\includegraphics[scale=0.4,trim=0cm 0cm -1cm 0cm,angle=270]{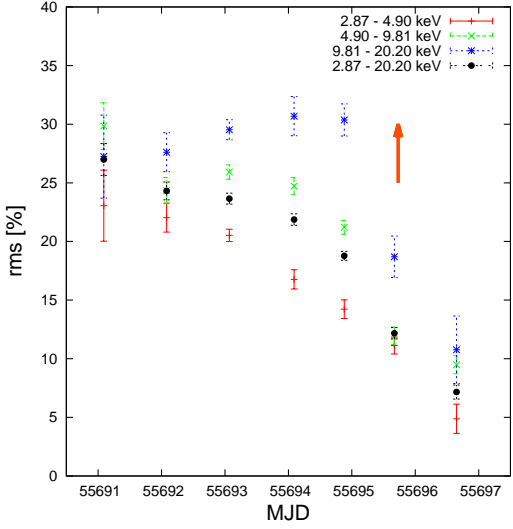}
\caption{Fractional integrated 1/128--10 Hz rms versus time in the first 7 observations for the bands considered. The orange arrow represents the time of the radio emission. All values are plotted with 1$\sigma$ error bars.}
\label{fig:rmsall}
\end{figure}

Only in the first 7 observations we detect QPOs ($Q>2$) and/or broad band components ($Q<2$) in at least some energy bands. We used the power spectrum of the fifth observation in band 0 (MJD 55694.884, Fig. \ref{fig:psex}) as a reference to identify four significant ($\sigma>3$, single-trial) components: a main QPO $L_{LF}$, its harmonic $L_{LF}^{+}$, a broad band noise component $L_b$, and another broad band component $L_b^{-}$ at lower frequency. In our analysis we reported all components with single trial significance $\sigma \ge 3$ and additionally those components with significance between 2$\sigma$ and 3$\sigma$ that could be identified as $L_{LF}$, $L_{LF}^{+}$, $L_b$, or $L_b^{-}$. Table \ref{tab:fmax} shows $\nu$, Q, rms, significance ($\sigma$) and reduced $\chi^2$ for every fitted component in the 7 observations analyzed (\#1--7) for all the energy bands. We also report the 99.87\% upper limits calculated fixing $\nu$ and Q to values equal to the most significant corresponding component between the energy bands fitted in the same observation. Empty lines mean that no components were fitted and no upper limit could be determined.\\
Figs. \ref{fig:fmax}$g$--$h$ show the frequencies of the fitted QPOs (triangles for $L_{LF}$, diamonds for $L_{LF}^{+}$), broad band components (squares for $L_b$, circles for $L_b^{-}$), and unidentified narrow ($Q>2$) components (pentagons), and their rms versus time in band 0, respectively. Solid symbols indicate significant components and open symbols components with significance between 2$\sigma$ and 3$\sigma$. The 2--3$\sigma$ unidentified component of observation \#6 (see Table \ref{tab:fmax}, bottom) is included in our plot because its characteristic frequency matches with the subharmonic frequency of the identified component $L_{LF}$. Similarly, two 2--3$\sigma$ unidentified components fitted in observation \#7 (Fig. \ref{fig:typeb}) were reported, as one matches with the subharmonic frequency of $L_{LF}$, and the other with $L_{b}$. Squares and circles were slightly shifted to the right for clarity.  \\
Always referring to band 0, in the first 5 observations one significant low frequency QPO ($L_{LF}$) was fitted for each spectrum and only the third observation shows a significant harmonic ($L_{LF}^{+}$). The $L_{LF}$ frequency increases with time from $\sim$ 1.1 Hz to $\sim$ 5.8 Hz while its rms decreases from $\sim$ 17$\%$ to $\sim$ 10$\%$ (see Table \ref{tab:fmax}). Observation \#7 shows a significant QPO with $\nu_{max}$ = 4.7 Hz (Fig. \ref{fig:typeb}). The peak characteristics ($\nu_{max}$ = 4.7 Hz, $Q$ = 9, rms $\sim$ 4.8) of and the low 1/128--10 Hz rms ($\sim$ 7.2\%) associated with this QPO, are characteristics of type-B QPOs (e.g. Casella et al. 2005). Considering also the 2--3$\sigma$ QPO fitted in observation \#6 ($\nu_{max}$ = 5.7 Hz, $\sigma$ $\sim$ 2.5), in observations \#6--7 $L_{LF}$ frequency and rms are not anti-correlated anymore. The characteristic frequency of $L_{LF}$ decreases from $\sim$ 5.8 Hz to $\sim$ 4.7 Hz while the rms still decreases from $\sim$ 6$\%$ to $\sim$ 5$\%$. \\
One significant broad band component ($L_b$) with $\nu_{max}$ in the interval $\sim$ 2--4 Hz was fitted in observations \#1--6. The rms of this component decreases with time (from 20$\%$ to 9$\%$), with a clear decreasing trend observable only in observations \#5--6, while its $\nu_{max}$ remains almost in the same frequency range (around 3 Hz). In observations \#5--7 we fitted another broad band component ($L_b^{-}$) characterized by an increasing $\nu_{max}$ (from observation \#5 to \#7) in the interval $\sim$ 0.06--0.66 Hz and rms between 2\% and 4\%.\\
The timing features in the other energy bands are reported in Fig. \ref{fig:fmax}$a$--$f$. Similarly to panels $g$--$h$, plots $a$--$b$, $c$--$d$, and $e$--$f$ show frequency and rms evolution for power spectral components fitted in bands 1, 2, and 3 respectively. No significant characteristic frequency shift was detected between energy bands in any power spectral component, while the rms values are systematically higher for higher energies (Table \ref{tab:fmax}). In band 1 $L_{LF}$ frequency increases with time (from $\sim$ 1.1 Hz to $\sim$ 6.5 Hz) in the first 6 observations while no-significant QPO was fitted in observation \#7. The behavior of $L_{LF}$ characteristic frequency in bands 2-3 is mostly identical to band 1 for observations \#1--5, but we observe some differences in observations \#6--7. The 2--3$\sigma$ QPO ($\sigma$ $\sim$ 2.2) fitted in observation \#6 (band 2) seems to break the anti-correlation between frequency and rms shown in observations \#1--5, but in band 3 the QPO frequency error bar is too big to infer any trend. However, the anti-correlation is evident in observation \#7, where a significant QPO was fitted in bands 2-3 with lower characteristic frequency compared to observation \#5. The rms of $L_{LF}$ in band 1 decreases as the QPO frequency increases, but in bands 2--3 this trend is progressively weaker. Indeed, in band 3 the $L_{LF}$ rms slightly oscillates around $\sim$ 17$\%$ and $\sim$ 11$\%$ in the first 5 observations and decreases to $\sim$ 11$\%$ only in the last 2 observations. \\
The broad band component ($L_b$) frequency slightly varies around $\sim$ 5 Hz in observations \#3--5 (band 1), while no significant broad band components were fitted in observations \#6 and \#7. In band 2 $L_b$ frequency shows a clear decreasing trend only in the last three observations ($\nu$ $\sim$ 4.4--1.6 Hz), while does not show any clear trend in band 3. $L_b$ rms decreases smoothly with time in band 1 (from $\sim$ 22$\%$ to $\sim$ 15$\%$), but it does not show the same trend in the other two energy bands. In band 2 we observe a clear decrease of $L_b$ rms only in observations \#5--6 (from $\sim$ 20$\%$ to $\sim$ 7$\%$ ) and in band 3 it oscillates between 22\% and 27\%.\\
Apart from the full energy, $L_b^{-}$ was fitted only in observation \#5 (band 1) and \#6 (band 1--3), but it is significant just at low photon energy (band 1, \#6). $L_b^{-}$ frequency and rms behavior in band 1 is mostly identical to band 0. \\
 
\section{Model fitting}
\label{sec:mod}

We fit the power spectra of the first 5 observations using \textsc{propfluc} (ID11; ID12; IK13). Whereas original explorations of the model (ID11; ID12) used computationally intensive Monte Carlo simulations, IK13 developed an exact analytic version of the model, allowing us for the first time to explore its capabilities systematically. We also investigate the relation between the values we obtained from the previously described phenomenological fitting of several Lorentzians (\S \ref{sec:obs} and \S \ref{sec:res}) and the model physical parameters (see Table 2 in ID12 for a  summary and description of all the physical parameters). \\ 

\subsection{The model}
\label{sec:modmod}

\begin{figure} 
\center
\includegraphics[scale=0.4,angle=270]{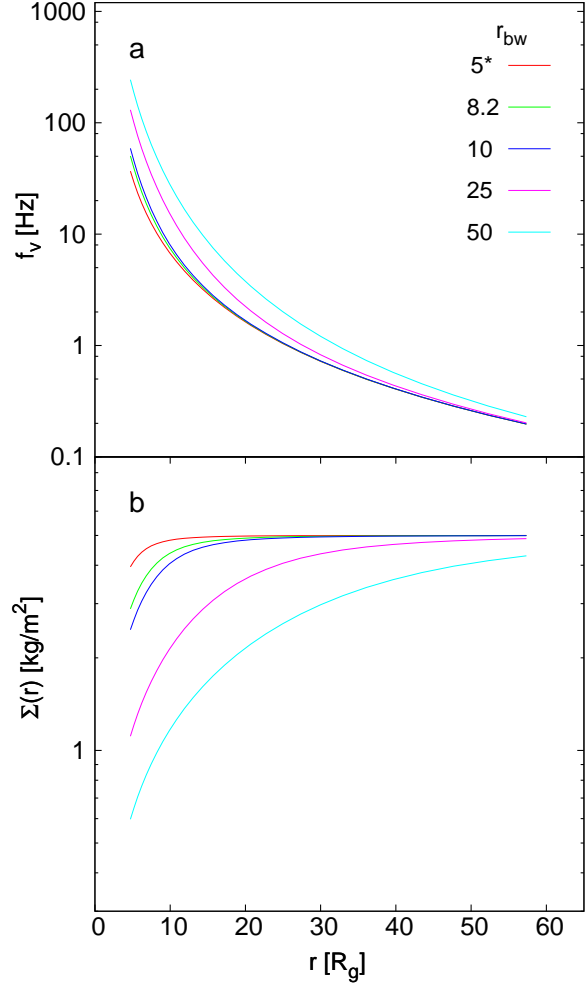}
\caption{Viscous frequency (a) and surface density profile (b) versus inner flow radius computed varying $r_{bw}$ (Fig. \ref{fig:simul}$b$).}
\label{fig:den}
\end{figure}

\textsc{propfluc} (\S \ref{sec:int}) parameterizes the flow surface density profile, which is required to calculate both the precession frequency and local viscous frequency, as a bending power law\footnote{ID12 showed that this surface density profile describes that measured from Fragile et al. (2007)'s simulations.}:
\begin{equation}
{\Sigma(r)}= \frac{\Sigma_0 \dot{M}_0}{cR_g}
\frac{x^\lambda}{(1+x^\kappa)^{(\zeta+\lambda)/\kappa}}
\label{eqn:sigma}
\end{equation}
where $x=r/r_{bw}$ and $r_{bw}$ is a break radius such that $\Sigma(r) \sim r^{-\zeta}$ for $r \gg r_{bw}$ and $\Sigma(r) \sim r^{-\lambda}$ for $r \ll r_{bw}$, with the sharpness of the break controlled by the parameter $\kappa$ (Fig. \ref{fig:den}b shows $\Sigma(r)$ examples for different $r_{bw}$ values). Here, $\dot{M}_0$ is the average mass accretion rate over the course of a single observation and $\Sigma_0$ is a dimensionless normalization constant. Throughout this paper, we employ the convention that $r \equiv R/R_g$ is radius expressed in units of $R_g$. The surface density drop off at the bending wave radius, $r_{bw}$, is due to the torque created by the radial dependence of LT precession ($\nu_{LT}$ $\propto$ $\sim r^{-3}$); i.e. essentially the inner regions try to precess quicker than the outer regions. Outside $r_{bw}$, bending waves (pressure waves) strongly couple the flow together but inside $r_{bw}$, material falls quickly towards the black hole (Lubow et al. 2002; Fragile et al. 2007). The precession frequency of the flow is given by (Liu \& Melia 2002):
\begin{equation}
\nu_{prec} = \nu_{qpo} = \frac{\int_{r_{i}}^{r_{o}} f_{LT}f_{k}\Sigma(r) r^3 dr}
{\int_{r_{i}}^{r_{o}} f_{k}\Sigma(r) r^3 dr},
\label{eqn:fprec}
\end{equation}
where $f_k$ is the Keplerian orbital frequency and $f_{LT}$ is the point particle Lense-Thirring (LT) precession frequency at $r$ (ID11). The bending wave radius carries information about the shape of the inner flow because it is dependent on the scaleheight factor of the flow, $h/r$: 
\begin{equation}
r_{bw} = 3(h/r)^{-4/5}a_{*}^{2/5}
\label{eqn:rbw}
\end{equation}
where $a_{*}$ is the dimensionless spin parameter.\\
If the mass is conserved on long timescales, the viscous frequency can be expressed as (Frank, King \& Raine 2002; ID12):
\begin{equation}
\nu_{visc}(r)= \frac{\dot{M}_0}{2 \pi R^2 \Sigma(r)}
\label{eqn:fv}
\end{equation}
\textsc{propfluc} assumes that the power spectrum of mass accretion rate fluctuations generated at $r$ is a zero-centered Lorentzian cutting off at the viscous frequency. The model splits the flow up into rings and assumes that a constant fractional variability, $F_{var}$, is generated per decade in radius. The resolution of the model is set by the number of rings per decade in radial extent, $N_{dec}$; i.e. the interval between $r=10$ and $r=100$ is split into $N_{dec}$ rings. Consequently, the fractional variability in the mass accretion rate, $\dot{M} (r,t)$, at each ring is $F_{var} /\sqrt{N_{dec}}$, so that $N_{dec}$ of these time series multiplied together has a fractional variability $F_{var}$. The emitted luminosity from a ring at $r_n$ is then assumed to be $\propto r_{n}^{2-\gamma}\Sigma(r_n) \dot{M}(r_n,t)$, where the emissivity index $\gamma>2$ is a model parameter, and the total emitted luminosity is simply the sum of the contributions from each ring. Thus, the low frequency break in the power spectrum corresponds roughly to $\nu_{visc}(r_o)$ (Churazov, Gilfanov \& Revnivtsev 2001; Ingram \& Done 2010). The high frequency break, however, does not correspond to $\nu_{visc}(r_i)$ because interference between radiation from different rings in the flow has a significant influence on the shape of the broad band noise at high frequencies (ID11; IK13). The emissivity index also clearly affects the shape of the high frequency noise: increasing $\gamma$ increases the amount of high frequency noise in the power spectrum. 

\subsection{Exploration of model parameters}
\label{sec:exp}

\begin{figure} 
\center
\includegraphics[scale=0.3,angle=270]{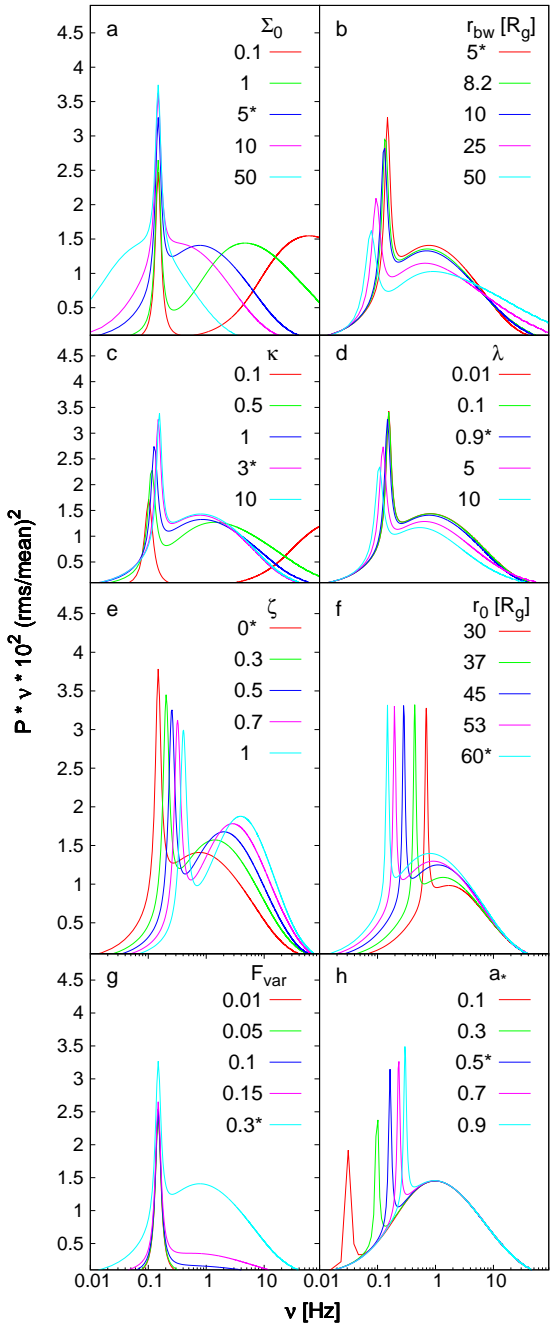}
\caption{Power spectra computed varying each of the main physical parameters of the model in turn as indicated. Asterisk indicates value of each parameter for all other computations.}
\label{fig:simul}
\end{figure}

To better understand the relation between the phenomenological parameters derived from Lorentzian power spectral fit characteristics and the physical parameters of \textsc{propfluc}, and to provide guidance in fitting this model to observed power spectra whose rough characteristics are known, we compute model power spectra with a Nyquist frequency of 128 Hz and vary one by one the main parameters. We fix the BH mass ($10 M_{\odot}$), the dimensionless spin parameter ($a_*=0.5$), the inner radius ($r_i=4.5 R_{g}$, so that $r_i>r_{ISCO}$), the bending wave radius ($r_{bw}=8.2$, so that $h/r \sim 0.2$), and the emissivity index ($\gamma=4.0$). We use a resolution $N_{dec}=25$ and include a QPO with fixed width and rms. The results of the calculations are shown in Fig. \ref{fig:simul}. Every plot specifies the parameter values, the asterisk in each panel denotes the value of that parameter used for all the other computations. \\
Fig. \ref{fig:simul}$a$ shows that the centroid frequency of the broad band component (hereafter $\nu_{b}$) decreases as $\Sigma_0$ increases, while the centroid frequency of the QPO (hereafter $\nu_{qpo}$) does not change. This can be understood from equations \ref{eqn:fprec} and \ref{eqn:fv}. Eq. \ref{eqn:fv} shows that increasing $\Sigma_0$ decreases $\nu_{visc} (r_0) \approx \nu_b$ but, since both integrals in Eq. \ref{eqn:fprec} contain the surface density, the constant $\Sigma_0$ cancels in the calculation of $\nu_{QPO}$. In contrast, the \textit{shape} of the surface density profile affects both the broad band noise and $\nu_{qpo}$. Eq. \ref{eqn:fprec} shows that the precession frequency of the entire flow is a surface density weighted average of $\nu_{LT}(r)$ (the precession frequency of a test mass a distance $r$ from the black hole). Therefore increasing the surface density at large $r$ slows down precession and increasing the surface density at small $r$ speeds up precession. Eq. \ref{eqn:fv} and Fig. \ref{fig:den}a--b show that increasing the surface density in any region decreases the viscous frequency in that region and increasing the \textit{gradient} of $\Sigma(r)$ increases the \textit{range} of frequencies at which the broad band noise contains significant power.\\
Fig. \ref{fig:simul}$b$ shows that $\nu_{qpo}$ decreases with increasing $r_{bw}$. This is because $r_{bw}$ governs where in the flow the surface density starts to drop off (see Fig. \ref{fig:den}b), and so increasing it weights the surface density towards the outer regions of the flow. Since, as demonstrated in Fig. \ref{fig:den}, increasing $r_{bw}$ slightly reduces the average surface density, this slightly increases the viscous frequency at the inner and outer rings in the flow, $\nu_{visc}(r_i)$ and $\nu_{visc}(r_o)$. This causes $\nu_b$ to increase by a small amount (which is difficult to see in Fig. \ref{fig:simul}b because of the QPO), and its affect on the high frequency power is complicated by the emissivity and interference between radiation from different regions of the flow (see IK13). Similar considerations are also valid for Fig. \ref{fig:simul}c--e.\\
Fig. \ref{fig:simul}f shows $\nu_{qpo}$ increasing as $r_o$ decreases, roughly following the trend $\nu_{qpo}\propto r_o^{-C}$, where $C$ is a positive constant ($C\sim 2.2$ for the fiducial model parameters). This can be understood if we assume a constant surface density profile (i.e. $\zeta=0$ and $r_{bw}$ is small) and use the weak field approximation for LT precession, $\nu_{LT}(r)$ $\propto r^{-3}$, to obtain a simplified version of Eq. 2 in Ingram, Done, \& Fragile (2009):
\begin{equation}
\nu_{qpo} = \frac{5 a_*}{\pi} \frac{ [ 1-(r_i/r_o)^{1/2} ] }{ r_o^{5/2} r_i^{1/2} [1-(r_i/r_o)^{5/2}] } \frac{c}{R_g}.
\end{equation}
We see that, for $r_i/r_{o} \ll 1$, the $r_o$ dependence of $\nu_{qpo}$ reduces to $\nu_{qpo}(r_o)\propto \sim r_o^{-5/2} \sim r_o^{-2.2}$. We also see from this equation that increasing $a_*$ increases the QPO frequency, as demonstrated in Fig. \ref{fig:simul}h. Thus, the parameters which most affect the QPO frequency are $r_o$ ($\sim$ quadratic) and $a_*$ (linear).

\subsection{The fit}
\label{sec:modfit}

\begin{figure} 
\center
\includegraphics[scale=0.4,angle=270, trim=40 150 20 140 ]{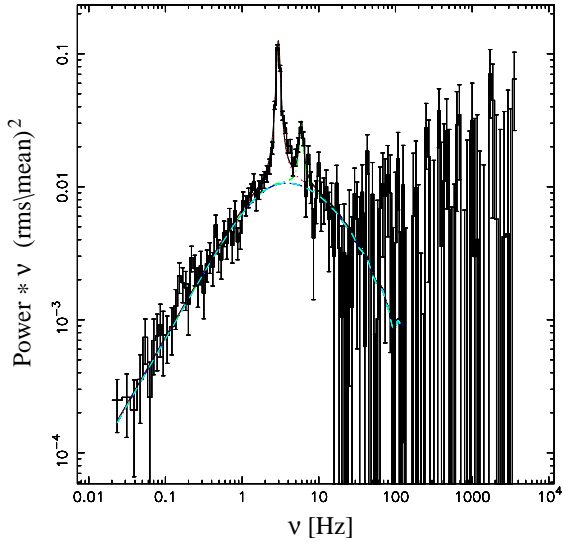}
\caption{\textsc{propfluc} fit of the third power spectrum.}
\label{fig:propex}
\end{figure}

\begin{figure} 
\center
\includegraphics[scale=0.4,angle=270]{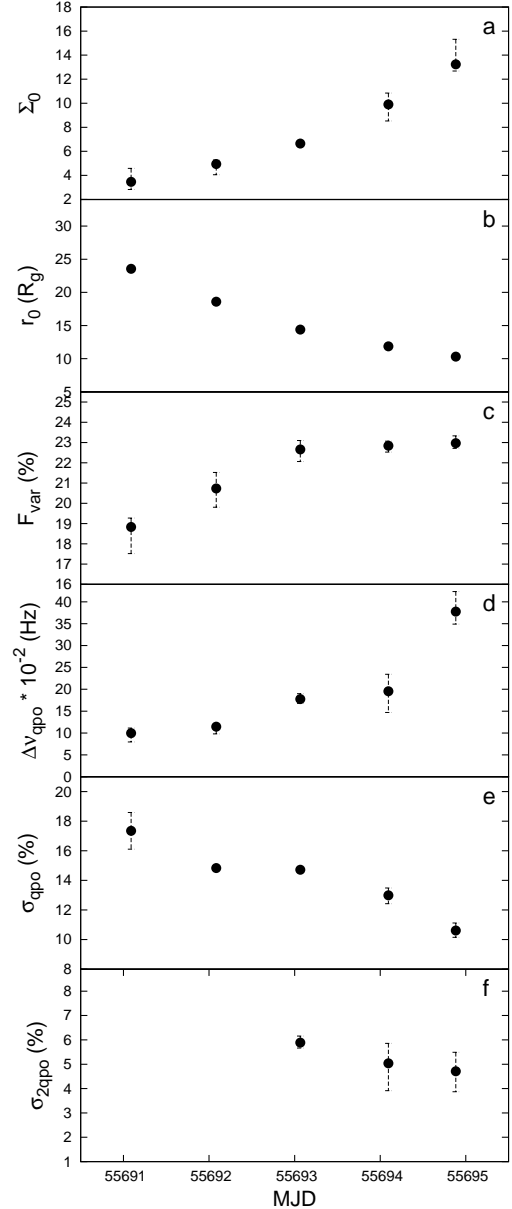}
\caption{\textsc{propfluc} best fit parameters versus time. All the points were plotted with 1$\sigma$ error bars.}
\label{fig:ingram}
\end{figure}

As described in \S\ref{sec:modmod}, the model is based on the inner accretion flow variability. For this reason, in order to avoid contamination from the disc, the best data choice for fitting would be observations in the high energy band ($\sim$ 10--20 keV), as mentioned in ID11. Because of the low count rate, we considered a wide 2.87--20.20 keV band. The model assumes that all the variability is generated in the hot flow, so formally, in this scenario, the only effect of the disc is to suppress the variability amplitudes at lower energies by dilution. Of course the possibility that the variability is generated in the disc and then propagates towards the inner flow affecting its emission, cannot be excluded, but for our first explorative fit we just considered the simple scenario described above. We note that, using the spectral model \textit{Tbabs * smedge * (discbb + nthComp + gauss)} (Wilms, Allen \& McCray 2000; Zdziarski, Johnson \& Magdziarz 1996; Zycki, Done \& Smith 1999; Mitsuda et al. 1984), we find that the disc contribution to the flux in band 1 increases from $\sim$ 38$\%$ in observation \#1 to $\sim$ 61$\%$ in observation \#7 (it contributes significantly less in the other bands). Since the rms in band 1 decreases from $\sim$ 33$\%$ to $\sim$ 5$\%$ for observations \#1-7, our assumption that the disc is stable implies that the fractional rms of the flow in this band is $\sim$ 53$\%$ to $\sim$ 13$\%$ for observation \#1-7 respectively, which is reasonable. \\
Since \textsc{propfluc} is not intended to explain the SIMS, we only fit the first 5 observations and leave a discussion of qualitative interpretation of observation \#6 and \#7 to \S \ref{sec:dis}, in the absence of statistically acceptable fits. We fitted logarithmically binned data points in the frequency range 1/128--128 Hz, using the same resolution for data and model. We used $N_{dec}$ = 15 for all the fits. Compared to power spectra computed using higher resolution (see \S \ref{sec:exp}) we did not observe any significant difference in $\chi^2$. The difference produced by a larger number of rings is appreciable only in the higher frequency region of the power spectra, where our data error bars are large. We combined the QPO with the broad band variability by multiplication instead of addition mode (the total flux is the product between the two types of variability instead of their sum, see IK13). Although our observations do not allow to differentiate between the multiplicative and the additive mode, the multiplicative mode is based on the more physically realistic scenario that the precession modulates the emission. Because of the inner flow precession, the brightest part of the inner flow moves in and out of the observer's line of sight and angles to the line of sight vary, causing variations in the projected area (IK13).\\
For all the fits we fixed the parameters $\zeta$, $\lambda$, $\kappa$, the bending wave radius $r_{bw}$, the emissivity $\gamma$, the mass M, and the dimensionless spin parameter of the black hole $a_{*}$. The free parameters are $\Sigma_{0}$, the truncation radius $r_{o}$, the fractional variability $F_{var}$, the fundamental QPO width $\Delta \nu$, and the rms of the fundamental and harmonic QPOs. \\
Fig. \ref{fig:propex} shows the \textsc{propfluc} fit of the third observation, Tab. \ref{tab:results} shows the best fit parameters, the main QPO frequency $\nu_{QPO}$, and the reduced $\chi^{2}_{\nu}$ for each of the 5 observations considered, and Fig. \ref{fig:ingram} shows the evolution of the physical parameters with time.\\
All parameter values show a clear trend. $\Sigma_{0}$ increases from $\sim$ 3.5 to $\sim$ 13 continuously. In the same way the truncation radius decreases from $\sim$ 24 $R_g$ to $\sim$ 10 $R_g$, indicating an average truncation radius recession speed of about 2 km/h. The fractional variability increases from $\sim$ 18$\%$ to $\sim$ 23$\%$ in the first 3 observations and shows no significant change beyond. The fundamental QPO width increases continuously with time while its rms decreases from $\sim 18\%$ to $\sim 10\%$.\\

\section{Discussion}
\label{sec:dis}

\begin{figure}
\center
\includegraphics[scale=0.35,angle=270, trim=40 150 60 140 ]{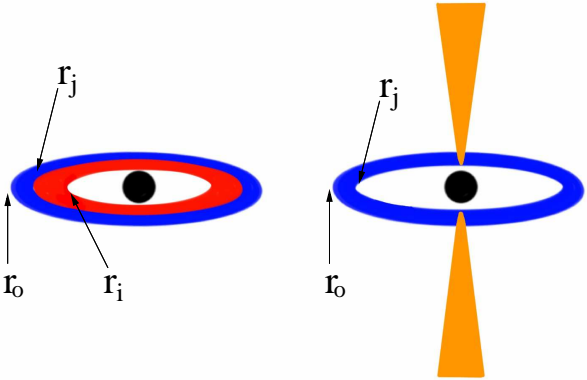}
\caption{Schematic representation of the transition between HS (a) and SIMS (b).}
\label{fig:jetm}
\end{figure}

As described by Stiele et al. (2012), the source follows the usual behavior of BHBs. Using the state classification described in Belloni (2010), the source either is in the HS or in the HIMS during the first 5 observations. Between observation \#5 and \#7 we observe rms dropping off (from $\sim$ 18$\%$ to $\sim$ 7$\%$) simultaneously to the detection of radio emission (Miller-Jones et al. 2011), and in observation \#7 we identify a significant (4.7$\sigma$, single trial) type-B QPO in the 2.87--20.20 keV energy band. This shows the source entered the SIMS between observation \#5 and \#7, a conclusion previously considered by Stiele et al. (2012), who however did not report the QPO in observation \#7.\\ 
We used the \textsc{profluc} model (ID11; ID12; IK13) for fitting the power spectra of the first 5 observations in the 2.87-20.20 keV band. As described in \S\ref{sec:int}, the model assumes that the variability generated by each region of the inner flow peaks at the local viscous time scale. This is contrary to results of General Relativistic Magneto-Hydrodynamic (GRMHD) simulations, which display variability peaking approximately at the local \textit{orbital} timescale (Dexter \& Fragile 2011; Armitage \& Reynolds 2003). These simulations, however, disagree with observations since BHBs display a high frequency break in their power spectra at $\nu \lesssim 10$ Hz, whereas simulations assuming a $10 M_\odot$ black hole exhibit variability up to a high frequency break of $\sim 100$ Hz, $\sim$1/3 the orbital frequency at $5 R_g$ (e.g., Fig. 10 of Dexter \& Fragile 2011). The same discrepancy with simulations is seen for Active Galactic Nuclei. For example, the power spectrum of the Seyfert 1 galaxy NGC 4051 displays a break at $\sim 8\times 10^{-4}$ Hz (Vaughan et al. 2011), whereas the orbital frequency at $5 R_g$ is $\sim 1.7\times 10^{-2}$ Hz (assuming a black hole mass of $\sim 1.7\times10^{6}M_\odot$). While it is clearly important that this inconsistency is addressed, we note that assuming the characteristic variability timescale to be orbital rather than viscous would require an inner flow radius of $r_i \sim 75 R_g$ in order to fit the observed power spectra in this paper. This strongly implies that considering the viscous timescale is more appropriate for black holes, even in light of evidence that pulsars show variability on the orbital timescale (Revnivtsev et al. 2009).\\
In order to better explore the possibilities of the model, we ran a series of computations varying its physical parameters. The model parameter mainly affecting QPO frequency is the truncation radius $r_o$, which sets the physical dimension of the precessing inner flow, responsible for the QPO generation. As can be seen in Fig. \ref{fig:simul}, most other parameters (but not $\Sigma_0$ and $F_{var}$) also affect the QPO frequency. Every parameter affects the broad band noise in a characteristic way. For example, varying the bending wave radius $r_{bw}$ we observe evident correlated variations in peak frequency and slope of the high frequency tail of the broad band component, varying $\zeta$ has almost no effect on the high frequency tail, but clearly changes its width and peak frequency, $\Sigma_0$ variations affect only the peak frequency. Variations in black hole spin $a_{*}$ have no effect at all on the broad band component. It is clear that in order to constrain all physical parameters of the model, very good counting statistics are needed to clearly define the precise shape of the broad band noise. \\
In our fits we fixed $\lambda$, $\zeta$, $\kappa$, $r_{bw}$, the emission coefficient $\gamma$, black hole mass $M$, and spin $a_{*}$. Fixing $\lambda$, $\zeta$, $\kappa$, and $r_{bw}$ implies fixing the surface density profile throughout the rising phase of the outburst. ID12 noticed that an evolution of $r_{bw}$ is expected when the truncation radius decreases, because when the inner flow is illuminated by an increasing number of disc photons, its electron temperature drops and hence its scaleheight factor, $h/r$, collapses. In our fit we fixed the bending wave radius because our spectra are too noisy above 10 Hz to measure it independently, so we are not able to eliminate the degeneracy between $r_o$ and $r_{bw}$ in determining the QPO frequency. For this reason we caution that the fit results in this work were obtained fixing the density profile of the inner flow (except for $r_o$), so that they must be interpreted with care. \\ 
The QPO frequency increase over observations \#1--5 corresponds to an $r_o$ decrease from $\sim$ 24 to $\sim$ 10 $R_g$. From spectral analysis, Stiele et al. (2012) report a constant truncation radius ($r_o$ $\sim$ 20-22 km) throughout the whole outburst without specifying uncertainties. As described before, the \textsc{profluc} physical parameter mainly affecting the QPO frequency is $r_o$, so that, in order to fit QPOs during the rising phase of the outburst, the truncation radius has to vary during this phase. Because of the data quality, RXTE spectral range, and the limitations of the model used by Stiele et al. (2012) (see Merloni et al. 2001), the spectral estimation of the inner radius is of limited use in the comparison with the \textsc{propfluc} results. \\
The surface density normalization constant $\Sigma_0$ increases from $\sim$ 3.5 to $\sim$ 13.2. For a given annulus in the inner flow, $\Sigma_0$ is proportional to surface density divided by mass accretion rate (Eq. \ref{eqn:sigma}). During the rising phase of the outburst the mass accretion rate increases with time, so the $\Sigma_0$ increase means that the surface density increases faster than the mass accretion rate, i.e. matter is accumulating in the flow during this phase of the outburst. This is consistent with the results of ID12 on XTE J1550-564. \\
The fractional variability $F_{var}$ shows a linear increasing trend in the first 3 observations and holds almost stable ($\sim$ 23$\%$) in the other observations. The fractional variability does not give us any detailed information about the physical mechanisms producing the variability, but it quantifies the turbulent nature of the accreting material per radial decade. ID12 show that $F_{var}$ increases continuously decreasing truncation radius, but we do not observe the same trend over all the observations, possibly because in our fit we fixed the bending wave radius.\\
As described, timing properties change in observations \#6 and \#7, compared to observations \#1--5, with simultaneous radio emission. The changes are more abrupt at higher photon energies (Fig. \ref{fig:rmsall}). The rms decreases from $\sim$ 18$\%$ to $\sim$ 7$\%$ and $L_{LF}$ frequency decreases as well, breaking the monotonically increasing trend of the first 5 observations. Observations \#6 and \#7 also show a prominent low frequency broad band component ($L_{b}^{-}$) that is not understandable in terms of the 2-component power spectra produced by \textsc{profluc}, which is the reason why we applied \textsc{profluc} only to the first 5 observations.  Belloni et al. (1997) consider emptying and replenishing cycles of the inner accretion disc, caused by viscous thermal instability, to explain variability on timescale of tens of minutes in  GRS 1915+105. Feroci et al. (1999) suggest material ejection to explain both spectral and timing properties of this same source, also in view of the correlation between the innermost disc temperature and the QPO frequency observed by Muno et al. (1999).  Similarly, assuming the truncation radius reaches its smallest value at maximum luminosity (and so maximum accretion rate), the $\nu_{QPO}$ decrease observed in our data can be explained by the depletion of inner flow material between $r_i$ and a certain radius $r_j$ ($r_o>r_j>r_i$) simultaneously to the radio emission. This scenario is shown schematically in Fig. \ref{fig:jetm}. This depletion can be caused by either ejection or increased accretion between $r_i$ and $r_j$. As a consequence, the surface density between $r_i$ and $r_j$ drops off, so that the high frequencies (corresponding to smaller radii) contribute less to the QPO frequency (Eq. \ref{eqn:fprec}) and $F_{var}$ (so the noise level) decreases.   \\
The low frequency broad band component $L_{b}^{-}$ might be explained in this scenario as the result of mass accretion rate fluctuations propagating from the disc towards the inner flow.\\

\section{Conclusions}

We analyzed the evolution of MAXI J1543-564 during its 2011 outburst identifying the transition between LHS/HIMS and SIMS, occurring between observation \#5 and \#6. Analyzing the source in different energy bands, we found that in this transition changes in rms are more evident at higher photon energy. Using the mass accretion rate fluctuation/precessing flow model \textsc{propfluc}, we provided a physical interpretation of the first 5 observations in terms of truncation radius, fractional variability, mass accretion rate, and surface density evolution. We suggest that the source behavior in observation \#6 and \#7, and so the transition between LHS and SIMS, might be caused by mass depletion in the innermost part of the accretion flow due to ejection and/or enhanced accretion associated with the simultaneous radio emission. This physical scenario is consistent with our timing analysis in different energy bands.

\begin{table*}
\caption{Multi- Lorentzian best fit parameters for observations 1-7 in 4 different energy bands.}\centering
\label{tab:fmax}
\begin{tabular}{cccccccc}

\hline
Date 	& Power Spectrum & Energy band & $\nu$ & Q & rms & $\sigma$ & {$\chi^2_{red}$} \\
{[MJD]} & Component & {[keV]} & {[Hz]} & & {[\%]} & & \\[1.0ex]
\hline\hline

 & & 2.87--4.90  & 1.07$^{+0.04}_{-0.05}$ & 4.17$^{+2.07}_{-1.01}$ & 17.97$^{+1.85}_{-1.85}$ & 4.86 & 1.03 \\[0.5ex]
 & & 4.90--9.81  & 1.03$^{+0.03}_{-0.02}$ & 3.42$^{+1.10}_{-0.84}$ & 20.80$^{+1.57}_{-1.64}$ & 6.36 & 1.07 \\[0.5ex]
 & & 9.81--20.20 & 1.04$^{+0.02}_{-0.02}$ & 8.38$^{+2.55}_{-2.55}$ & 17.42$^{+2.07}_{-2.07}$ & 4.21 & 0.81 \\[-2.1ex]
\raisebox{3.5ex}{55691.089(\#1)} & \raisebox{3.5ex}{L$_{LF}$} & 2.87--20.20 & 1.06$^{+0.02}_{-0.02}$ & 
5.26$^{+1.34}_{-1.05}$ & 17.48$^{+1.27}_{-1.24}$ & 7.05 & 0.88 \\[2.5ex]

 & & 2.87--4.90  & 1.72$^{+0.02}_{-0.03}$ & 6.54$^{+3.07}_{-1.64}$  & 13.37$^{+1.11}_{-1.06}$ & 6.32 & 1.17 \\[0.5ex]
 & & 4.90--9.81  & 1.75$^{+0.02}_{-0.01}$ & 10.91$^{+7.32}_{-2.63}$ & 14.34$^{+0.92}_{-0.89}$ & 8.04 & 0.87 \\[0.5ex]
 & & 9.81--20.20 & 1.74$^{+0.02}_{-0.02}$ & 6.46$^{+2.68}_{-1.48}$  & 18.69$^{+1.57}_{-1.47}$ & 6.35 & 0.83 \\[-2.1ex]
\raisebox{3.5ex}{55692.084(\#2)} & \raisebox{3.5ex}{L$_{LF}$} & 2.87--20.20 & 1.75$^{+0.01}_{-0.01}$ & 
7.81$^{+1.55}_{-1.15}$ & 14.74$^{+0.67}_{-0.66}$ & 11.17 & 0.91 \\[2.5ex]

 & & 2.87--4.90  & 2.97$^{+0.02}_{-0.02}$ & 7.82$^{+1.61}_{-1.14}$ & 11.25$^{+0.57}_{-0.54}$ & 10.49 & 1.04 \\[0.5ex]
 & & 4.90--9.81  & 2.97$^{+0.01}_{-0.01}$ & 9.51$^{+1.28}_{-0.98}$ & 16.06$^{+0.43}_{-0.43}$ & 18.93 & 1.04 \\[0.5ex]
 & & 9.81--20.20 & 2.97$^{+0.01}_{-0.02}$ & 11.06$^{+4.25}_{-2.07}$& 16.37$^{+0.72}_{-0.67}$ & 12.18 & 0.88 \\[-2.1ex]
\raisebox{3.5ex}{55693.066(\#3)} & \raisebox{3.5ex}{L$_{LF}$} & 2.87--20.20 & 2.97$^{+0.01}_{-0.01}$ & 
8.60$^{+0.69}_{-0.60}$ & 14.60$^{+0.29}_{-0.29}$ & 25.14 & 1.22 \\[2.5ex]

 & & 2.87--4.90  & 4.39$^{+0.03}_{-0.07}$ & 12.35$^{+5.33}_{-5.33}$ & 8.37$^{+1.06}_{-0.86}$ & 4.86 & 0.84 \\[0.5ex]
 & & 4.90--9.81  & 4.42$^{+0.05}_{-0.05}$ & 9.18$^{+1.44}_{-1.15}$ & 14.57$^{+1.00}_{-1.00}$ & 7.30 & 0.89 \\[0.5ex]
 & & 9.81--20.20 & 4.46$^{+0.07}_{-0.05}$ & 7.00$^{+1.57}_{-1.10}$ & 19.62$^{+1.55}_{-1.60}$ & 6.14 & 1.00 \\[-2.1ex]
\raisebox{3.5ex}{55694.095(\#4)} & \raisebox{3.5ex}{L$_{LF}$} & 2.87--20.20 & 4.381$^{+0.02}_{-0.02}$ & 
11.27$^{+3.19}_{-1.87}$ & 13.01$^{+0.51}_{-0.49}$ & 13.18 & 0.91  \\[2.5ex]

 & & 2.87--4.90  & 5.67$^{+0.15}_{-0.14}$ & 6.61$^{+9.13}_{-2.64}$ & 7.35$^{+1.65}_{-1.26}$ & 2.93 & 1.04 \\[0.5ex]
 & & 4.90--9.81  & 5.84$^{+0.10}_{-0.11}$ & 8.48$^{+2.52}_{-1.32}$ & 11.10$^{+1.04}_{-1.01}$ & 5.50 & 1.07 \\[0.5ex]
 & & 9.81--20.20 & 5.83$^{+0.11}_{-0.12}$ & 6,52$^{+2.95}_{-1.83}$ & 17.41$^{+2.66}_{-2.44}$ & 3.57 & 0.91 \\[-2.1ex]
\raisebox{3.5ex}{55694.884(\#5)} & \raisebox{3.5ex}{L$_{LF}$} & 2.87--20.20 & 5.77$^{+0.05}_{-0.05}$ & 
7.98$^{+1.20}_{-0.93}$ & 10.43$^{+0.61}_{-0.61}$ & 8.53 & 0.80  \\[2.5ex]

 & & 2.87--4.90  & 6.52$^{+0.42}_{-0.76}$ & 3.69$^{+7.40}_{-1.87}$ & 5.51$^{+1.64}_{-1.26}$ & 2.19 & 1.10 \\[0.5ex]
 & & 4.90--9.81  & 5.74$^{+0.22}_{-0.22}$ & 4.96$^{+4.99}_{-2.37}$ & 6.08$^{+1.42}_{-1.12}$ & 2.15 & 1.08 \\[0.5ex]
 & & 9.81--20.20 & 6.19$^{+0.27}_{-0.36}$ & 4.45$^{+8.57}_{-2.31}$ & 12.84$^{+2.87}_{-2.67}$& 2.40 & 1.08 \\[-2.1ex]
\raisebox{3.5ex}{55695.669(\#6)} & \raisebox{3.5ex}{L$_{LF}$} & 2.87--20.20 & 5.75$^{+0.41}_{-0.49}$ & 1.81$^{+1.18}_{-0.86}$ & 
6.35$^{+2.05}_{-1.28}$ & 2.49 & 1.08 \\[2.5ex]

 & & 2.87--4.90  & 4.69 & 9.00 & $<$ 5.32 & - & - \\[0.5ex]
 & & 4.90--9.81  & 4.83$^{+0.10}_{-0.06}$ & 10.38$^{+4.50}_{-4.60}$ & 5.34$^{+0.91}_{-0.67}$ & 3.97 & 0.84 \\[0.5ex]
 & & 9.81--20.20 & 4.86$^{+0.06}_{-0.17}$ & 10.49$^{+4.62}_{-4.62}$ & 11.37$^{+1.59}_{-1.42}$& 4.00 & 1.21 \\[-2.1ex]
\raisebox{3.5ex}{55696.650(\#7)} & \raisebox{3.5ex}{L$_{LF}$} & 2.87--20.20 & 4.69$^{+0.05}_{-0.04}$ & 
9.00 & 4.76$^{+0.36}_{-0.37}$ & 6.48 & 1.01  \\[2.5ex]


\hline

 & & 2.87--4.90  & 2.28 & 4.55 & $<$ 6.48 & - & - \\[0.5ex]
 & & 4.90--9.81  & 2.28$^{+0.56}_{-0.09}$ & 4.55$^{+5.92}_{-3.71}$ & 13.05$^{+7.56}_{-2.57}$ & 2.53 & 1.07\\[0.5ex]
 & & 9.81--20.20 & 2.99$^{+ 0.17}_{-0.14}$ & 4.55 & $<$ 22.34 & 14.73$^{+3.19}_{-3.19}$ & 0.81\\[-2.1ex]
\raisebox{3.5ex}{55691.089(\#1)} & \raisebox{3.5ex}{L$_{LF}^{+}$} & 2.87--20.20 & 2.17$^{+0.15}_{-0.14}$ & 6.78 &
 5.42$^{+2.10}_{-2.12}$ & 1.29 & 0.88 \\[2.5ex]

 & & 2.87--4.90  & 4.03$^{+0.21}_{-0.13}$ & 5.18$^{+4.74}_{-2.32}$ & 10.31$^{+2,34}_{-1.86}$ & 2.78 & 1.17  \\[0.5ex]
 & & 4.90--9.81  & 4.03 & 5.18 & $<$  12.69 & - & - \\[0.5ex]
 & & 9.81--20.20 & - & - & - & - & \\[-2.1ex]
\raisebox{3.5ex}{55692.084(\#2)} & \raisebox{3.5ex}{L$_{LF}^{+}$} & 2.87--20.20 & 4.03 & 5.18 & $<$ 8.00 & - & - \\[2.5ex]

 & & 2.87--4.90  & 5.87$^{+0.16}_{-0.15}$ & 5.18$^{+4.74}_{-2.32}$ & 6.24$^{+1.27}_{-1.25}$ & 2.50 & 1.04 \\[0.5ex]
 & & 4.90--9.81  & 5.87$^{+0.16}_{-0.15}$ & 7.29$^{+1.88}_{-1.88}$ & 6.57$^{+1.15}_{-1.02}$ & 3.23 & 1.04 \\[0.5ex]
 & & 9.81--20.20 & - & - & - & - & - \\[-2.1ex]
\raisebox{3.5ex}{55693.066(\#3)} & \raisebox{3.5ex}{L$_{LF}^{+}$} & 2.87--20.20 & 5.85$^{+0.11}_{-0.11}$ & 
6.55$^{+2.89}_{-1.53}$ & 6.16$^{+0.83}_{-0.73}$ & 4.23 & 1.22 \\[2.5ex]

 & & 2.87--4.90  & 8.86 & 6.69 & $<$ 9.50 & - & - \\[0.5ex]
 & & 4.90--9.81  & 8.86 & 5.88 & 6.87$^{+3.06}_{-2.21}$ & 1.55 & 0.89 \\[0.5ex]
 & & 9.81--20.20 & - & - & - & - & - \\[-2.1ex]
\raisebox{3.5ex}{55694.095(\#4)} & \raisebox{3.5ex}{L$_{LF}^{+}$} & 2.87--20.20 & 8.86$^{+0.26}_{-0.28}$ & 
6.69$^{+3.02}_{-3.02}$ & 5.88$^{+1.60}_{-1.16}$ & 2.55 & 0.91 \\[2.5ex]

\end{tabular}
\end{table*}

\begin{table*}
\begin{tabular}{cccccccc}
\hline

Date 	& Power Spectrum & Energy band & $\nu$ & Q & rms & $\sigma$ & {$\chi^2_{red}$} \\
{[MJD]} & Component & {[keV]} & {[Hz]} & & {[\%]} & & \\[1ex]


\hline\hline

 & & 2.87--4.90  & - & - & - & - & - \\[0.5ex]
 & & 4.90--9.81  & 3.30 & 0 & $<$ 30.16 & - & - \\[0.5ex]
 & & 9.81--20.20 & - & - & - & - & - \\[-2.1ex]
\raisebox{3.5ex}{55691.089(\#1)} & \raisebox{3.5ex}{L$_{b}$} & 2.87--20.20 & 3.30$^{+1.33}_{-0.87}$ & 
0 & 19.55$^{+2.37}_{-2.32}$ & 4.22 & 0.88 \\[2.5ex]

 & & 2.87--4.90  & - & - & - & - & - \\[0.5ex]
 & & 4.90--9.81  & 3.641$^{+0.86}_{-0.67}$ & 0 & 22.12$^{+1.71}_{-1.67}$ & 6.64 & 0.87 \\[0.5ex]
 & & 9.81--20.20 & 7.881$^{+7.40}_{-3.33}$ & 0 & 24.65$^{+4.60}_{-3.73}$ & 3.31 & 0.83 \\[-2.1ex]
\raisebox{3.5ex}{55692.084(\#2)} & \raisebox{3.5ex}{L$_{b}$} & 2.87--20.20 & 3.61$^{+0.41}_{-0.36}$ & 
0 & 21.38$^{+0.89}_{-0.89}$& 11.96 & 0.91  \\[2.5ex]

 & & 2.87--4.90  & 4.813$^{+0.74}_{-0.69}$ & 0 & 18.71$^{+1.42}_{-1.21}$ & 7.72 & 1.04 \\[0.5ex]
 & & 4.90--9.81  & 3.755$^{+0.52}_{-0.50}$ & 0 & 21.25$^{+1.25}_{-1.27}$ & 8.38 & 1.04 \\[0.5ex]
 & & 9.81--20.20 & 4.261$^{+0.66}_{-0.63}$ & 0 & 27.40$^{+1.67}_{-1.69}$ & 8.09 & 0.88 \\[-2.1ex]
\raisebox{3.5ex}{55693.066(\#3)} & \raisebox{3.5ex}{L$_{b}$} & 2.87--20.20 & 3.77$^{+0.27}_{-0.27}$ & 
0 & 19.98$^{+0.61}_{-0.63}$ & 15.83 & 1.22 \\[2.5ex]

 & & 2.87--4.90  & 4.64$^{+0.74}_{-0.69}$ & 0 & 15.14$^{+1.74}_{-1.69}$ & 4.48 & 0.83 \\[0.5ex]
 & & 4.90--9.81  & 3.44$^{+0.78}_{-0.69}$ & 0 & 20.28$^{+1.78}_{-1.68}$ & 6.02 & 0.89 \\[0.5ex]
 & & 9.81--20.20 & 1.71$^{+0.84}_{-0.49}$ & 0 & 21.61$^{+3.04}_{-2.47}$ & 4.38 & 1.00 \\[-2.1ex]
\raisebox{3.5ex}{55694.095(\#4)} & \raisebox{3.5ex}{L$_{b}$} & 2.87--20.20 & 3.42$^{+0.52}_{-0.49}$ & 
0 & 18.19$^{+1.70}_{-1.05}$ & 8.69 & 0.91 \\[2.5ex]

 & & 2.87--4.90  & 5.36$^{+2.01}_{-1.63}$ & 0 & 13.18$^{+1.92}_{-1.87}$ & 3.52 & 1.04 \\[0.5ex]
 & & 4.90--9.81  & 4.36$^{+0.80}_{-0.72}$ & 0 & 20.28$^{+1.28}_{-1.24}$ & 8.11 & 1.07 \\[0.5ex]
 & & 9.81--20.20 & 3.54$^{+1.34}_{-0.98}$ & 0 & 25.89$^{+3.45}_{-2.95}$ & 4.39 & 0.91 \\[-2.1ex]
\raisebox{3.5ex}{55694.884(\#5)} & \raisebox{3.5ex}{L$_{b}$} & 2.87--20.20 & 3.69$^{+0.67}_{-0.57}$ & 
0 & 16.90$^{+0.99}_{-0.93}$ & 8.53 & 0.80 \\[2.5ex]

 & & 2.87--4.90  & - & - & - & - & - \\[0.5ex]
 & & 4.90--9.81  & 2.50$^{+0.23}_{-0.22}$ & 1.69$^{+0.89}_{-0.70}$ & 6.94$^{+1.17}_{-1.00}$ & 3.48 & 1.08 \\[0.5ex]
 & & 9.81--20.20 & - & - & - & - & - \\[-2.1ex]
\raisebox{3.5ex}{55695.669(\#6)} & \raisebox{3.5ex}{L$_{b}$} & 2.87--20.20 & 3.08$^{+1.04}_{-0.91}$ & 
0 & 8.56$^{+1.42}_{-1.38}$ & 3.10 &  1.08 \\[2.5ex]

 & & 2.87--4.90  & 1.55 & 0 & $<$ 5.03 & - & - \\[0.5ex]
 & & 4.90--9.81  & 1.55$^{+0.94}_{-0.56}$ & 0 & 6.68$^{+1.25}_{-1.01}$ & 3.32 & 0.84 \\[0.5ex]
 & & 9.81--20.20 & - & - & - & - & - \\[-2.1ex]
\raisebox{3.5ex}{55696.650(\#7)} & \raisebox{3.5ex}{L$_{b}$} & 2.87--20.20 & - & 
- & - &  - & -  \\[2.5ex]

\hline

 & & 2.87--4.90  & 0.06$^{+0.05}_{-0.03}$ & 0 & 3.27$^{+0.84}_{-0.76}$ & 2.15 & 1.04 \\[0.5ex]
 & & 4.90--9.81  & 0.06 & 0.13 & $<$ 5.07 & - & - \\[0.5ex]
 & & 9.81--20.20 & - & - & - & - & - \\[-2.1ex]
\raisebox{3.5ex}{55694.884(\#5)} & \raisebox{3.5ex}{L$_{b}^{-}$} & 2.87--20.20 & 0.057$^{+0.05}_{-0.02}$ & 
0.13$^{+0.48}_{-0.48}$ & 3.54$^{+0.90}_{-0.58}$ & 3.04 & 0.80  \\[2.5ex]

 & & 2.87--4.90  & 0.19$^{+0.06}_{-0.05}$ & 0 & 4.76$^{+0.56}_{-0.53}$ & 4.49 & 1.10 \\[0.5ex]
 & & 4.90--9.81  & 0.26$^{+0.76}_{-0.11}$ & 0 & 4.62$^{+0.92}_{-0.81}$ & 2.86 & 1.08 \\[0.5ex]
 & & 9.81--20.20 & 0.52$^{+2.04}_{-0.26}$ & 0 & 10.47$^{+6.41}_{-1.98}$& 2.64 & 1.08\\[-2.1ex]
\raisebox{3.5ex}{55695.669(\#6)} & \raisebox{3.5ex}{L$_{b}^{-}$} & 2.87--20.20 & 0.13$^{+0.04}_{-0.03}$ & 
0 & 4.42$^{+0.47}_{-0.44}$ & 5.00 & 1.08  \\[2.5ex]

 & & 2.87--4.90  & - & - & - & - & - \\[0.5ex]
 & & 4.90--9.81  & - & - & - & - & - \\[0.5ex]
 & & 9.81--20.20 & - & - & - & - & - \\[-2.1ex]
\raisebox{3.5ex}{55696.650(\#7)} & \raisebox{3.5ex}{L$_{b}^{-}$} & 2.87--20.20 & 0.27$^{+0.39}_{-0.15}$ & 
0 & 2.49$^{+0.79}_{-0.56}$ & 2.23 & 0.91 \\[2.5ex]

\end{tabular}
\end{table*}

\begin{table*}
\begin{tabular}{cccccccc}
\hline

Date 	& Power Spectrum & Energy band & $\nu$ & Q & rms & $\sigma$ & {$\chi^2_{red}$} \\
{[MJD]} & Component & {[keV]} & {[Hz]} & & {[\%]} & & \\
\hline\hline

 &  &  & 0.561$^{+0.01}_{-0.04}$ & 6.88$^{+18.05}_{-3.11}$ & 6.65$^{+1.17}_{-0.97}$ & 3.42 & 1.10\\[-1ex]
\raisebox{1.5ex} {55692.084(\#2)}& \raisebox{1.5ex}{?}&\raisebox{1.5ex}{2.87--4.90}& 2.74$^{+0.06}_{-0.09}$ &
6.10$^{+11.11}_{-2.33}$ & 10.28$^{+1.94}_{-1.63}$ & 3.16 & 1.10 \\[2.5ex]

 &  & {2.87--4.90} & 3.38$^{+0.27}_{-0.22}$ & 2.40$^{+1.61}_{-0.89}$ & 6.57$^{+1.21}_{-1.15}$ & 2.85 & 1.10\\[-1ex]
\raisebox{1.5ex} {55695.669(\#6)}& \raisebox{1.5ex}{$L_{LF}$ sub?}&{2.87--20.20}& 3.11$^{+0.12}_{-0.09}$ & 6.33$^{+3.45}_{-3.45}$ & 3.67$^{+1.42}_{-0.83}$ & 2.20 & 1.10\\[2.5ex]

 & $L_{LF}$ sub? & {2.87--20.20} & 2.43$^{+0.06}_{-0.04}$ & 9.00 & 2.79$^{+0.44}_{-0.49}$ & 2.85 & 1.01\\[-1ex]
\raisebox{1.5ex} {55696.650(\#7)}& $L_{b}$ ? &{2.87--20.20}& 1.57$^{+0.19}_{-0.12}$ & 3.27$^{+4.16}_{-1.63}$ & 2.70$^{+0.78}_{-0.64}$ & 2.23 & 1.01\\[2.5ex]

\end{tabular}
\end{table*}

\begin{table*}
\caption{Best \textsc{propfluc} fit physical parameters for observations 1-5. A $\sim$ symbol indicates
that the parameter has been fixed.}\centering
\label{tab:results}
\begin{tabular}{l|c|c|c|c|c} 

\hline
Observations & 1 & 2 & 3 & 4 & 5 \\
\hline
\hline

$\Sigma_0$ & 3.47$^{+1.12}_{-0.64}$ & 4.95$^{+0.34}_{-0.89}$ & 6.65$^{+0.25}_{-0.04}$ & 9.90$^{+0.94}_{-1.37}$ & 13.24$^{+2.08}_{-0.56}$ \\ [1.5ex]
$F_{var}$ [\%] & 18.84$^{+0.44}_{-1.32}$ & 20.73$^{+0.79}_{-0.92}$ & 22.66$^{+0.44}_{-0.60}$ & 22.84$^{+0.23}_{-0.31}$ & 22.97$^{+0.36}_{-0.26}$\\[1.5ex]
$\zeta$	  & 0    & $\sim$ & $\sim$ & $\sim$ & $\sim$\\ [1.5ex]
$\lambda$ & 0.90 & $\sim$ & $\sim$ & $\sim$ & $\sim$\\ [1.5ex]
$\kappa$  & 3.00 & $\sim$ & $\sim$ & $\sim$ & $\sim$\\ [1.5ex]
$r_i$	  & 4.50 & $\sim$ & $\sim$ & $\sim$ & $\sim$\\ [1.5ex]
$r_{bw}$  & 8.24 & $\sim$ & $\sim$ & $\sim$ & $\sim$\\ [1.5ex]
$r_o$ & 23.56$^{+0.05}_{-0.07}$ & 18.59$^{+0.01}_{-0.01}$ & 14.40$^{+0.00}_{-0.00}$ & 11.87$^{+0.00}_{-0.01}$ & 10.32$^{+0.01}_{-0.01}$\\ [1.5ex]
$\Delta \nu_{qpo}(10^{-2})$ & 9.97$^{+1.17}_{-2.01}$ & 11.44$^{+0.61}_{-1.64}$ & 17.75$^{+1.23}_{-0.98}$ & 19.55$^{+3.88}_{-4.86}$	& 37.76$^{+4.60}_{-2.84}$\\ [1.5ex]
$\sigma_{qpo}$ [\%] & 17.34$^{+1.24}_{-1.24}$ & 14.83$^{+0.08}_{-0.18}$	& 14.71$^{+0.08}_{-0.09}$ & 12.99$^{+0.50}_{-0.57}$ & 10.61$^{+0.51}_{-0.46}$\\ [1.5ex]
$\sigma_{2apo}$	[\%]& $0.$ & $0.$ & 5.88$^{+0.27}_{-0.22}$ & 5.04$^{+0.82}_{-0.12}$ & 4.71$^{+0.78}_{-0.84}$\\ [1.5ex]
$\gamma$ & $ 4.0$ & $\sim$ & $\sim$ & $\sim$ & $\sim$ \\ [1.5ex]
$M (M_{\odot})$	& $10.0$ & $\sim$ & $\sim$ & $\sim$ & $\sim$\\ [1.5ex]
$a$ & $ 0.5$ & $\sim$ & $\sim$ & $\sim$ & $\sim$ \\ [1.5ex]
$\nu_{qpo}$ [Hz] & 1.06 & 1.75 & 2.97 & 4.38 & 5.78 \\ [1.5ex]
$\chi^{2}_{\nu}$ & $0.90$ & $0.91$ & $1.25$ & $0.89$ & $0.99$\\ [1.5ex]
\\ 
 
\end{tabular}
\end{table*}

\label{lastpage}

\begin{thebibliography}{99}

\bibitem[\protect\citeauthoryear{Ar{\'e}valo 
\& Uttley}{2006}]{2006MNRAS.367..801A} Ar{\'e}valo P., Uttley P., 2006, MNRAS, 367, 801 

\bibitem[Armitage 
\& Reynolds(2003)]{2003MNRAS.341.1041A} Armitage, P.~J., \& Reynolds, C.~S.\ 2003, MNRAS, 341, 1041 

\bibitem[Belloni et al.(1997)]{1997ApJ...488L.109B} Belloni, T., Mendez, 
M., King, A.~R., van der Klis, M., 
\& van Paradijs, J.\ 1997, ApJ, 488, L109 

\bibitem[\protect\citeauthoryear{Belloni, Psaltis, 
\& van der Klis}{2002}]{2002ApJ...572..392B} Belloni T., Psaltis D.,
van der Klis M., 2002, ApJ, 572, 392 

\bibitem[Belloni et 
al.(2005)]{2005A&A...440..207B} Belloni, T., Homan, J., Casella, P., et al.\ 2005, A\&A, 440, 207 

\bibitem[Belloni(2010)]{2010LNP...794...53B} Belloni, T.~M.\ 2010, LNP, Berlin Springer Verlag, 794, 53 

\bibitem[Churazov et al.(2001)]{2001MNRAS.321..759C} Churazov, E., 
Gilfanov, M., \& Revnivtsev, M.\ 2001, MNRAS, 321, 759 

\bibitem[Dexter 
\& Fragile(2011)]{2011ApJ...730...36D} Dexter, J., \& Fragile, P.~C.\ 2011, ApJ, 730, 36 

\bibitem[\protect\citeauthoryear{Done, Gierli{\'n}ski, 
\& Kubota}{2007}]{2007A&ARv..15....1D} Done C., Gierli{\'n}ski M., Kubota A.,
2007, A\&ARv, 15, 1 

\bibitem[\protect\citeauthoryear{Esin, McClintock, 
\& Narayan}{1997}]{1997ApJ...489..865E} Esin A.~A., McClintock J.~E., Narayan R., 1997, ApJ, 489, 865 

\bibitem[Fender et al.(2004)]{2004MNRAS.355.1105F} Fender, R.~P., Belloni, 
T.~M., \& Gallo, E.\ 2004, MNRAS, 355, 1105 

\bibitem[Fender et 
al.(2005)]{2005Ap&SS.300....1F} Fender, R., Belloni, T., \& Gallo, E.\ 2005, Ap\&SS, 300, 1

\bibitem[Feroci et 
al.(1999)]{1999A&A...351..985F} Feroci, M., Matt, G., Pooley, G., et al.\ 1999, A\&A, 351, 985 

\bibitem[\protect\citeauthoryear{Fragile et 
al.}{2007}]{2007ApJ...668..417F} Fragile P.~C., Blaes O.~M., Anninos P., 
Salmonson J.~D., 2007, ApJ, 668, 417 

\bibitem[Frank et al.(2002)]{2002apa..book.....F} Frank, J., King, A., 
\& Raine, D.~J.\ 2002, Accretion Power in Astrophysics, by Juhan Frank and Andrew King and Derek Raine, pp.~398.~ISBN 0521620538.~Cambridge, UK: Cambridge University Press, February 2002.,  

\bibitem[\protect\citeauthoryear{Gilfanov}{2010}]{2010LNP...794...17G} 
Gilfanov M., 2010, LNP, 794, 17

\bibitem[\protect\citeauthoryear{Ingram, Done, 
\& Fragile}{2009}]{2009MNRAS.397L.101I} Ingram A., Done C., Fragile P.~C., 2009,
MNRAS, 397, L101

\bibitem[\protect\citeauthoryear{Ingram 
\& Done}{2010}]{2010MNRAS.405.2447I} Ingram A., Done C., 2010, MNRAS, 405, 2447 

\bibitem[\protect\citeauthoryear{Ingram 
\& Done}{2011}]{2011MNRAS.415.2323I} Ingram A., Done C., 2011, MNRAS, 415, 2323 

\bibitem[Ingram 
\& Done(2012)]{2012MNRAS.419.2369I} Ingram, A., \& Done, C.\ 2012, MNRAS, 419, 2369 

\bibitem[Ingram 
\& Klis(2013)]{2013MNRAS.434.1476I} Ingram, A., \& Klis, M.~v.~d.\ 2013, MNRAS, 434, 1476

\bibitem[Jahoda et al.(1996)]{1996SPIE.2808...59J} Jahoda, K., Swank, 
J.~H., Giles, A.~B., et al.\ 1996, Proc. SPIE, 2808, 59 

\bibitem[Klein Wolt(2004)]{2004PhDT.......415K} Klein Wolt, M.\ 2004, 
Ph.D.~Thesis

\bibitem[Liu 
\& Melia(2002)]{2002ApJ...573L..23L} Liu, S., \& Melia, F.\ 2002, ApJ, 573, L23 

\bibitem[Lubow et al.(2002)]{2002MNRAS.337..706L} Lubow, S.~H., Ogilvie, 
G.~I., \& Pringle, J.~E.\ 2002, MNRAS, 337, 706 

\bibitem[\protect\citeauthoryear{Lyubarskii}{1997}]{1997MNRAS.292..679L} 
Lyubarskii Y.~E., 1997, MNRAS, 292, 679 

\bibitem[Matsuoka et al.(2009)]{2009PASJ...61..999M} Matsuoka, M., 
Kawasaki, K., Ueno, S., et al.\ 2009, PASJ, 61, 999 

\bibitem[Merloni et al.(2001)]{2001AIPC..599..770M} Merloni, A., Fabian, 
A.~C., 
\& Ross, R.~R.\ 2001, X-ray Astronomy: Stellar Endpoints, AGN, and the Diffuse X-ray Background, 599, 770 

\bibitem[Miller-Jones et al.(2011)]{2011ATel.3364....1M} Miller-Jones, 
J.~C.~A., Tzioumis, A.~K., Jonker, P.~G., et al.\ 2011, The Astronomer's 
Telegram, 3364, 1 

\bibitem[\protect\citeauthoryear{Mitsuda et
al.}{1984}]{1984PASJ...36..741M} Mitsuda K., et al., 1984, PASJ, 36, 741

\bibitem[Muno et al.(1999)]{1999ApJ...527..321M} Muno, M.~P., Morgan, 
E.~H., \& Remillard, R.~A.\ 1999, ApJ, 527, 321 

\bibitem[Negoro et al.(2011)]{2011ATel.3330....1N} Negoro, H., Nakahira, 
S., Ueda, Y., et al.\ 2011, The Astronomer's Telegram, 3330, 1 

\bibitem[\protect\citeauthoryear{Psaltis, Belloni, 
\& van der Klis}{1999}]{1999ApJ...520..262P} Psaltis D., Belloni T.,
van der Klis M., 1999, ApJ, 520, 262 

\bibitem[Remillard et al.(2002)]{2002ApJ...580.1030R} Remillard, R.~A., 
Muno, M.~P., McClintock, J.~E., \& Orosz, J.~A.\ 2002, ApJ, 580, 1030 

\bibitem[Remillard 
\& McClintock(2006)]{2006ARA&A..44...49R} Remillard, R.~A., \& McClintock, J.~E.\ 2006, ARA\&A, 44, 49 

\bibitem[Revnivtsev et 
al.(2009)]{2009A&A...507.1211R} Revnivtsev, M., Churazov, E., Postnov, K., \& Tsygankov, S.\ 2009, A\&A, 507, 1211

\bibitem[\protect\citeauthoryear{Shakura 
\& Sunyaev}{1973}]{1973A&A....24..337S} Shakura N.~I., Sunyaev R.~A., 1973,
A\&A, 24, 337 

\bibitem[Stiele et al.(2012)]{2012MNRAS.422..679S} Stiele, H., 
Mu{\~n}oz-Darias, T., Motta, S., \& Belloni, T.~M.\ 2012, MNRAS, 422, 679 

\bibitem[Sunyaev 
\& Truemper(1979)]{1979Natur.279..506S} Sunyaev, R.~A., \& Truemper, J.\ 1979, Nature, 279, 506 

\bibitem[Thorne 
\& Price(1975)]{1975ApJ...195L.101T} Thorne, K.~S., \& Price, R.~H.\ 1975, ApJ, 195, L101 

\bibitem[Uttley 
\& McHardy(2001)]{2001MNRAS.323L..26U} Uttley, P., \& McHardy, I.~M.\ 2001, MNRAS, 323, L26 

\bibitem[Uttley et al.(2005)]{2005MNRAS.359..345U} Uttley, P., McHardy, 
I.~M., \& Vaughan, S.\ 2005, MNRAS, 359, 345 

\bibitem[Vaughan et al.(2011)]{2011MNRAS.413.2489V} Vaughan, S., Uttley, 
P., Pounds, K.~A., Nandra, K., 
\& Strohmayer, T.~E.\ 2011, MNRAS, 413, 2489 

\bibitem[Wijnands 
\& van der Klis(1998)]{1998ApJ...507L..63W} Wijnands, R., \& van der Klis, M.\ 1998, ApJ, 507, L63 

\bibitem[Wijnands et al.(1999)]{1999ApJ...526L..33W} Wijnands, R., Homan, 
J., \& van der Klis, M.\ 1999, ApJ, 526, L33 

\bibitem[Wilms et al.(2000)]{2000ApJ...542..914W} Wilms, J., Allen, A., 
\& McCray, R.\ 2000, ApJ, 542, 914 

\bibitem[\protect\citeauthoryear{Zdziarski, Johnson, 
\& Magdziarz}{1996}]{1996MNRAS.283..193Z} Zdziarski A.~A., Johnson W.~N., Magdziarz P., 1996, MNRAS, 283, 193 

\bibitem[Zhang et al.(1995)]{1995ApJ...449..930Z} Zhang, W., Jahoda, K., 
Swank, J.~H., Morgan, E.~H., \& Giles, A.~B.\ 1995, ApJ, 449, 930

\bibitem[\protect\citeauthoryear{{\.Z}ycki, Done, 
\& Smith}{1999}]{1999MNRAS.309..561Z} {\.Z}ycki P.~T., Done C., Smith D.~A., 1999, MNRAS, 309, 561

\end{thebibliography}
\end{document}